\documentclass[preprint,showpacs,preprintnumbers,amsmath,amssymb]{revtex4}

\usepackage{graphicx}
\usepackage{dcolumn}
\usepackage{bm}
\usepackage{times}

\begin{document}

\title{Memory-Based Boolean game and self-organized phenomena on networks}

\author{HUANG Zi-Gang, WU Zhi-Xi, GUAN Jian-Yue, and WANG Ying-Hai}

\email{yhwang@lzu.edu.cn}

\affiliation{Institute of Theoretical Physics, Lanzhou University,
Lanzhou Gansu 730000, China}

\date{\today}

\begin{abstract}
We study a memory-based Boolean game (MBBG) taking place on the
regular ring, wherein each agent acts according to its local
optimal states of the last $M$ time steps recorded in memory, and
the agents in the minority are rewarded. One free parameter $p$
among $0$ and $1$ is introduced to denote the strength of the
agents' willing to make a decision according to its memory. We
find that, given proper willing strength $p$, the MBBG system can
spontaneously evolve to a state of better performance than the
random game; while for larger $p$, the herd behavior emerges which
reduces the system profit. By analyzing the dependence of the
system's dynamics on the memory capacity $M$, we find that a
higher memory capacity favors the emergence of the better
performance state, and effectively restrains the herd behavior,
therefore increases the system profit. Considering the high cost
of long-time memory, the enhancement of memory capacity for
restraining the herd behavior is also discussed, and the $M=5$ is
suggested to be one good choice.
\end{abstract}


\maketitle

Dynamical systems with many elements under mutual regulation or
influence, such as the systems naturally arise in biology
\cite{Kauffman} and in the social sciences \cite{Arthur_1},
underlie much of the phenomena associated with complexity. The
perspective of complex adaptive systems (CAS) composed of agents
under mutual influence have been proposed for understanding the
rich and complex dynamics of these real-life systems
\cite{Kauffman,Levin,Anderson,Yangcx}.

One of the simplest examples of a complex dynamical system is the
minority game \cite{ChalletPA} (MG) introduced by Challet and
Zhang as a simplification of Arthur's El Farol Bar attendance
problem \cite{Arthur_2}. Agents in the MG are designed to make
choice ($1$ or $0$, i.e. to attend a bar or refrain) based on the
aggregate signal (the global information in memory), i.e., which
value was in the majority for the last several time steps. The
agents in the minority are rewarded, and those in the majority
punished since resources are limited. The MG model can serve as a
general paradigm for resource allocation and load balancing in
multiagent systems and was study extensively
\cite{Quan,Yang,Challet,see}. In contrast to this mean-field
description of the MG, the Boolean game (BG) on the network of
interconnections between the agents was introduced in Ref.
\cite{Paczuski} considering that the agent can also respond to the
detailed information it receives from other specified agents. It
was established that coordination still arises out of local
interactions in the BG, and the system as a whole achieves
``better than random'' performance in terms of the utilization of
resources \cite{Galstyan,Chen,ZhouBoolean,Ma}. This contributes to
the solution of one basic question in studies of complexity, that
is, how large systems with only local information available to the
agents may become complex through a self-organized dynamical
process \cite{Paczuski}.

Many real-life systems often seem a black box to us: the outcome
may be observed, but the underlying mechanism is not visible. Herd
behavior, which describes the condition that many agents display
the same action, is one of the outcomes always present in
ecosystems while the corresponding mechanisms are unaware. The
herd behavior has been extensively studied in Behavioral Finance
and is found to be one factor of the origins of complexity that
may enhance the fluctuation and reduce the system profit
\cite{Eguluz,Xie,Lee,Wang,ZhouPL}. Also, the underlying mechanism
of the herd behavior is an interesting issue which has attracted
economists' and physicists' interests. Considering that herd
behavior still occur although the agents prefer to be in the
minority in some real-life cases, one should seek the mechanism of
the herd behavior from some other aspects rather than the agents'
willing to be in majority \cite{ZhouBoolean}.

In the previous studies of the BG, each agent acts according to
the Boolean function, i.e., gets its input from some other agents,
and maps the input to a state it will adopt in the subsequent
round \cite{Galstyan,Chen}. Inspired by the MG, we argue that the
agents should make decisions based on the knowledge of the past
records, and the historical memory of individuals plays a key role
in the evolutionary games. In the present work, we study a
memory-based Boolean game (MBBG) in which each agent modifies its
state based on its past experiences gained from the local
interaction with neighbors, and the agents in the minority of the
whole system are rewarded.

The global information is not available, and the agents also do
not know who are winners in the previous rounds. They can only
make use of the local information and gain experiences from local
interaction. It is worthwhile to emphasize that the agent's
ignorance of who are global winners is one of the main differences
from the previous studies on MG. Due to the lack of the global
information, each agent in our model attempts to be in the
minority in its own small region which consists of its immediate
neighbors and itself, considering that there should exist positive
correlation between being the minority in the whole system and in
its own local region. We call this ``local optimal assumption''
(LOA) of the agent system. Then, our model can be depicted as: in
the lack of the global information and in the belief of the LOA,
the agent pins its hope for winning in the whole system on the
effort to act as minority in its own region based on the local
experiences stored in memory.

Let us introduce the rules of the evolutionary MBBG. To simplify,
each agent is confined to a site of a regular network which is a
one-dimensional lattice with periodic boundary conditions and
coordination number $z=3$ \cite{Newman}. A local region for each
agent thus contain $7$ agents. When a round of game is over, each
agent will have the state information ($+1$ or $-1$) of its
neighbors. Then the agents are designed to know its \textbf{local
optimal state} (LOS) in the past round by means of
self-questioning, i.e., each agent adopts its anti-state to play a
virtual game with all its neighbors, and calculates the virtual
result. Comparing the virtual result with the actual one, each
agent gets its LOS which may turn it into minority of its own
local region. In condition that the counterbalance of the groups
with $+1$ and $-1$ appears in one agent's neighbors, its optimal
state is randomly set either as $+1$ or $-1$, because whichever
state the agent chooses, it will break the counterbalance and
compel its own side into majority. Then, the agent records the LOS
into memory. Taking into account the bounded capacity of the
agents, we assume that the agents are quite limited in their power
and can only retain the last $M$ LOS in memory. We would like to
reiterate that the so called ``local optimal state'' does not mean
the agent will be rewarded if has adopted it. Only the agents in
the global minority are rewarded by $1$, and therefore the system
profit equals to the number of agents in the global minority. This
is a main difference of our model from the Local Minority Game
\cite{Moelbert}.

There might be variability of the agents' belief of the LOA and
the willing to make decision based on records in memory. We define
the willing strength $p$ to add this effect into our model. That
in detail is, at each time step, each agent acts based on its
memory at probability $p$, or acts all by himself at probability
$1-p$. In the former case, the probability of making a decision
(choosing $+1$ or $-1$) for each agent depends on the ratio of the
numbers of $+1$ and $-1$ stored in its memory, i.e., the agent
chooses $+1$ with the probability
$P(+)=n_{+}/(n_{+}+n_{-})=n_{+}/{M}$ and $-1$ with the probability
$P(-)=1-P(+)$, where $n_{+}$ and $n_{-}$ are the numbers of $+1$
and $-1$, respectively. In the latter case, the agent simply
inherits its action in the last time step or chooses the opposite
action at a small probability $m$, named the mutation probability.
Following the previous work \cite{ZhouBoolean}, we set it to be
$0.01$. The introduction of $m$ adds several impulsive and
unstable ingredients to our model in view of the presence of the
irrational effect .

\begin{figure}
\includegraphics[width=8cm]{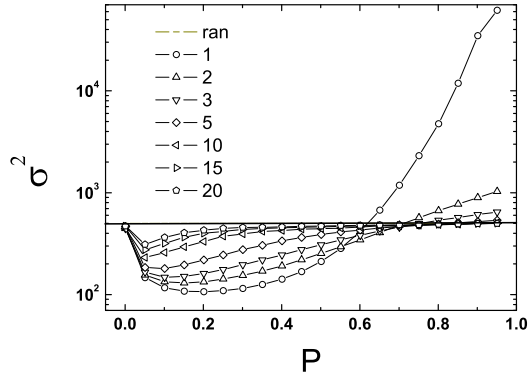}
\caption{The variance of the number of agents choosing $+1$ as a
function of willing strength $p$ with several different memory
capacities on the regular ring of size $2001$. The solid line
represents the system profit of the random choice game which
corresponds to the $M=0$ case. The system performs better than the
random game when $p$ is less than the intersection point
$p_{inter}^{M}$.}\label{fig:l}
\end{figure}

Simulations are carried out for a population of $N=2001$ agents
located on network sites. The time length $T=10^{4}$ are fixed. In
the initial state, $+1$ and $-1$ are uniformly distributed among
all the agents, and the memory information of each agent is
randomly assigned. We have checked that this assignment has no
contributions to the evolutionary behavior of the system. All the
simulation results presented in this paper are average of $50$
randomly assigned initial states.

The variance of systems
$\sigma^{2}=(1/T)\sum^{T}_{t=1}(A_{t}-N/2)^{2}$
\cite{ZhouBoolean,Ma} which is the cumulative standard deviation
from the optimal resource utilization over time length $T$, can be
considered as a global measure of the system's optimality. The
smaller $\sigma^{2}$ corresponds to better optimality of the
system and the more system profit. Here, $A_{t}$ denotes the
number of agents choosing $+1$ at time $t$. The simulation results
$\sigma^{2}$ as a function of $p$ with different memory capacity
$M$ are presented in Fig. \ref{fig:l}. The result of the random
game which is same as the MBBG with $M=0$ is also plotted for
comparison. In the random choice game, $A_{t}$ does not depend on
the previous states of the system, and its expectation is always
$\langle{A_{t}}\rangle=N/2$. The distribution of $A_{t}$ has a
Gaussian profile with the variance to the expectation $N/2$ as
$\sigma^{2}=0.25N$ in the limit of large $N$. For the MBBG with
$M\neq{0}$, it is noticeable that these systems can perform better
than that with random choice game when $p$ is in a certain
interval (see Fig. \ref{fig:l} the interval where
$\sigma^{2}<0.25N$). This is an evidence of the existence of a
self-organized process in the systems. At larger $p$, the herd
behaviors occur and the subsequent oscillations cause the greater
variances $\sigma^{2}$ than that of the random choice game. The
intersection points of the curves of the MBBG and that of the
random game (at $p=p_{inter}^{M}$, $M=1,2,...,20$) denote the same
system performance of them.

Let us firstly consider the extreme case $p=0$ which means that
the agents act all by themselves without considering the
historical memory. In this case, each agent merely changes its
action with the mutation probability $m$, and there is no
preferential choice for $+1$ and $-1$ so that no herd behaviors
occur. Following Ref. \cite{ZhouBoolean}, the expectation of
$A_{t+1}$ is,
\begin{equation}
 \langle{A_{t+1}}\rangle=A_{t}(1-m)+m(N-A_{t})=A_{t}+m(N-2A_{t}).
 \end{equation}
Assuming $A_{t}>N/2$, for $0<m<1/2$, we have $A_{t}>A_{t+1}>N/2$.
Thus, if large event has taken place initially in the system
(e.g., $A_{t=0}\gg{N/2}$, or $A_{t=0}\ll{N/2}$), the effort of $m$
will make $A_{t}$ slowly revert to the equilibrium position $N/2$.
It is easy to prove that even when the mutation probability $m$ is
very small, the system profit will be equal to random choice game
on condition that the evolutionary time $T$ is sufficiently long.
The simulation results for $p=0$ (see Fig. \ref{fig:l}) are in
well agreement with the our analysis.

The other extreme case is $p=1$ where the herd behavior prevails.
Comparing this case to the $p=0$ case, we can say that the
occurrence of the herd behavior is intimately related to the
mechanism of the memory-based actions. In this case, if the agents
choosing $+1$ and $-1$ are equally mixed up in the networks, then
the number of agents who record $+1$ as the LOS by
self-questioning (denoted by $S^{opt}_{+1,t}$) has the expectation
\begin{equation}
\langle{S^{opt}_{+1,t}}\rangle\doteq{N-A_{t}}.
\end{equation}
Thus, all over the system the collection of the agents' newly
recorded LOS is close to the anti-state of the present system. For
the system with small memory capacity $M$, e.g. $M=1$, the agents'
new states for the subsequent round $t+1$ gained from the records
in memory are actually their optimal states of the latest round,
and thus the expectation of $A_{t+1}$ is
\begin{equation}
\langle{A_{t+1}}\rangle=\langle{S^{opt}_{+1,t}}\rangle\doteq{N-A_{t}},
\end{equation}
with departure $|\langle{A_{t+1}}\rangle- N/2|\doteq|A_{t}-N/2|$.
One can see that the departure from $N/2$ does not reduce in
average, while the state of the winning side reverses. Therefore,
the prevalence of the herd behavior which is denoted by the large
durative oscillation will occur when $p=1$ and $M=1$. On the other
hand, for the system with larger values of the memory capacity
$M$, the agents have also stored more previous information in
memory besides the latest LOS. Based on more information their
state updates will not be so intense and irrational as that with
$M=1$. As a result, the behavior of the systems are mended by the
rationality of their agents. It is clear in Fig. \ref{fig:l} that
the oscillation of the system with larger $M$ is less acute than
that with smaller $M$ in the $p=1$ case. Furthermore, in the cases
of $p\in(p_{inter}^{M},1)$, it can also be found that the high
memory capacity of the agent can effectively restrain the herd
behavior and thus increase the system profit.

The existence of the self-organization demonstrated in Fig.
\ref{fig:l} can be understood by the dynamics of the system in the
mentioned two extreme cases: The action of the agent based on
memory with probability $p$ will induce oscillation, while the
independent mechanism with probability $1-p$ will lead to a long
reversion process to the equilibrium position $N/2$. Thus at a
proper value of $p$, the system can quickly arrive at the
equilibrium position after the occurrence of a large event, which
leads to more system profit than the random game. Also, we can see
that the underlying mechanism of the herd behavior is related to
the strength of the agents' willing of making decision based on
the historical memory.

\begin{figure}
\includegraphics[width=8cm]{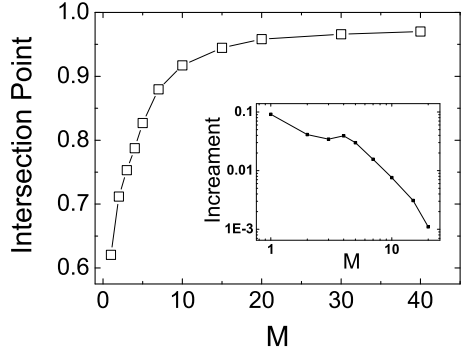}
\caption{The intersection point $p_{inter}^{M}$ of the random game
and the MBBG with different memory capacity $M$, for regular ring
with $N=2001$. The inset is the log-log plot of the increments of
$p_{inter}^{M}$ as a function of memory capacity
$M$.}\label{fig:2}
\end{figure}

In the following, we discuss the effect of the memory capacity $M$
to the behavior of the system in detail from two points, the
intersection $p_{inter}^{M}$ and the corresponding transition
rate, which will be defined in the following.

We have known that, the MBBG system can perform either better or
worse than the random game when the value of $p$ is smaller or
larger than $p_{inter}^{M}$. The case of the better performance is
due to the emergence of the agents' self-organization, and the
case of the worse performance is due to the prevalence of the herd
behavior. The relation between $p_{inter}^{M}$ and the memory
capacity $M$ are plotted in Fig. \ref{fig:2}. It is revealed that,
the region $(0,p_{inter}^{M})$ where system performs better than
random game broadens with the memory capacity $M$. That is to say,
the system with larger memory capacity has more probability of
self-organizing to the better-performance case. In addition, the
inset in Fig. \ref{fig:2} presents the increments of the
intersection point $\Delta{p}_{inter}^{M}$ when $M$ increases by
one (\emph{i.e.}
$\Delta{p}_{inter}^{M}=p_{inter}^{M+1}-p_{inter}^{M}$) as a
function of $M$. This measure corresponds to the ``marginal
return'' in economics. We can see that, when the memory capacity
is large, the increment is small. The scaling behavior at large
$M$ implies that $p_{inter}^{M}$ is arriving at a level number
close to $1$. It is remarkable that the behavior of the
$\Delta{p}_{inter}^{M}$ with $M$ is not monotonic. There exists
the special point at $M=4$ which implies that the $p_{inter}^{M}$
with $M=5$ is larger comparing to the value estimated from the
trend exhibited from all the other values of $M$.

\begin{figure}
\includegraphics[width=8cm]{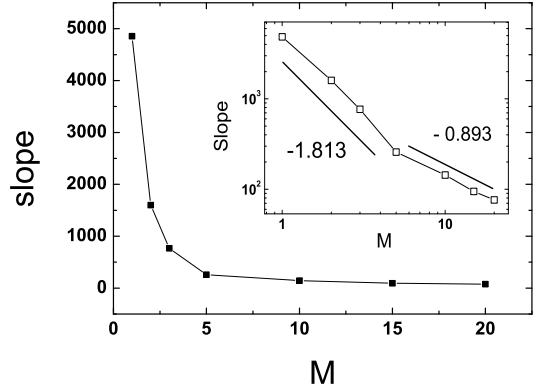}
\caption{The memory capacity $M$ and the slope of the variance at
the intersection point $p_{inter}^{M}$. The inset is the
corresponding log-log plot, where the crossover at $M=5$ is
obvious.}\label{fig:3}
\end{figure}

Inspired by the fact that in many situations the agents have to
operate in dynamic (and in general, stochastic) environments, we
can imagine that, due to some external impacts the willing
strength $p$ of the agents in our model may be not fixed, but vary
with time. In the case that $p$ fluctuates around $p_{inter}^{M}$,
there exists the transition from the case of the better
performance to the case where the herd behavior seriously impacts
the system profit. Let us now focus on the rate of the transition
between the two cases when $p$ is fluctuating.  For convenience,
we call this rate the ``transition rate'' which is different from
its traditional meaning in the study of the phase transition. It
is noticeable in Fig \ref{fig:l} that, at the intersection point
$p_{inter}^{M}$, different memory lengthes $M$ correspond to
different values of slope. We study the relation between the slope
at $p_{inter}^{M}$ (i.e. the transition rate) and the memory
capacity $M$ (see Fig. \ref{fig:3}). It can easily be found that,
the shorter the memory is, the rapider the transition from the two
cases would be. One can also consider the transition rate as a
measure of the system's risk of suffering from the herd behavior.
The results in Fig. \ref{fig:3} thus is the dependence between the
system's risk and the memory capacity $M$. It is clear that, those
systems with higher memory capacity can constrain the occurrence
of the herd behavior more efficiently.

On the other hand, if the question facing us is to bring down the
system's risk or to design a system with low risk of suffering
from the herd behavior, enhancing the agents' memory capacity is
indeed an effective way. However, the enhancement of the memory
capacity in real-life cases would cost much. In this context, it
is necessary to discuss how large memory capacity would be proper.
Interestingly, we find that the dependence of the slope on $M$
approximately obeys scaling laws with two exponents (the inset in
Fig. \ref{fig:3}). That is, for small $M$ ($M\leq5$) the scaling
exponent is about $-1.813$, after which, at larger $M$, there is a
crossover to $-0.893$. This behavior implies that, when the memory
length $M$ is already $5$ or larger, if increase $M$, the risk
reduces slower than the small $M$ cases. Also it is obvious that
the value of the transition rate at $M=5$ is already small. When
$M>5$ the effort to increase memory capacity which costs much can
not gain good mark in reducing system crisis. Further simulation
results show that the two-exponent scaling behavior divided by
$M=5$ holds for different system size $N$, coordination number $z$
and mutation probability $m$. Moreover, in the previous
intersection analysis we have proved that the $p_{inter}^{M}$ at
$M=5$ is comparatively large. Thus we argue that, $M=5$ may be a
good choice to improve the performance of the system.

In summary, inspired by the minority game, we studied a
memory-based Boolean Game on regular ring. The simulation results
with various memory capacity $M$ are discussed. We found that,
those systems with nonzero $M$ can perform better than that of the
random choice game when willing strength $p$ is in a certain
interval. This is reasonable evidence of the existence of a
self-organized process taking place within the agent system,
although only local information is available to the agents. The
memory capacity $M$ are found to have remarkable effect on the
agent system. That is, the larger the $M$ is, the more probably
the self-organized process would emerge since the value of
$p_{inter}^{M}$ increases. Moreover, larger memory capacity
corresponds to smaller degree of the herd behavior at large $p$,
and less risk of the system suffering from the herd behavior when
$p$ fluctuates around $p_{inter}^{M}$. In addition, we propose the
question of designing the system which is robust to the impact of
the herd behavior, and the choice of $M$ is also discussed
considering the high cost of enhancing $M$ in real-life cases.

We thank Dr. Xin-Jian Xu for helpful discussions and valuable
suggestions. This work was supported by the Fundamental Research
Fund for Physics and Mathematics of Lanzhou University under Grant
No. Lzu05008.


\begin{references}

\bibitem{Kauffman}
Kauffman S A 1993 \emph{The Origins of Order} (New York: Oxford
University Press)

\bibitem{Arthur_1}
Arthur W B 1999 \emph{Science} \textbf{284} 107

\bibitem{Levin}
Levin S A 1998 \emph{Ecosystems} \textbf{1} 431

\bibitem{Anderson}
Anderson P W, Arrow K and Pines D 1988 \emph{The Economy as an
Evolving Complex System} (Redwood City, CA: Addison-Wesley)

\bibitem{Yangcx}
Yang C X, Zhou T, Zhou P L, Liu J and Tang Z N 2005 \emph{Chin.
Phys. Lett.} \textbf{22} 1014

\bibitem{ChalletPA}
Challet D and Zhang Y C 1997 \emph{Physica} A \textbf{246} 407

\bibitem{Arthur_2}
Arthur W B 1994 \emph{Am. Econ. Assoc. Papers Proc.} \textbf{84}
406

\bibitem{Quan}
Quan H J, Wang B H, Hui P M and Luo X S 2001 \emph{Chin. Phys.
Lett.} \textbf{18} 1156

\bibitem{Yang}
Yang W S, Wang B H, Quan H J and Hu C K 2003 \emph{Chin. Phys.
Lett.} \textbf{20} 1659

\bibitem{Challet}
Challet D and Marsili M 1999 \emph{Phys. Rev.} E \textbf{60} R6271

\bibitem{see}
See http://www. unifr. ch/econophysics/minority/ for an extensive
collection of papers and references

\bibitem{Paczuski}
Paczuski M, Bassler K E and Corral \'{A} 2000 \emph{Phys. Rev.
Lett.} \textbf{84} 3185

\bibitem{Galstyan}
Galstyan A and Lerman K 2002 \emph{Phys. Rev.} E \textbf{66}
015103(R)

\bibitem{Chen}
Yuan B S, Wang B H and Chen K \emph{cond-mat}/0411664

\bibitem{ZhouBoolean}
Zhou T, Wang B H, Zhou P L, Yang C X and Liu J 2005 \emph{Phys.
Rev.} E \textbf{72} 046139

\bibitem{Ma}
Ma J, Zhou P L, Zhou T, Bai W J, Cai S M \emph{physics}/0604066

\bibitem{azquez}
V\'{a}zquez A 2000 \emph{Phys. Rev.} E \textbf{62} 4497

\bibitem{Eguluz}
Eguluz V M and Zimmermann M G 2000 \emph{Phys. Rev. Lett.}
\textbf{85} 5659

\bibitem{Xie}
Xie Y B, Wang B H, Hu B and Zhou T 2005 \emph{Phys. Rev.} E
\textbf{71} 046135

\bibitem{Lee}
Lee S and Kim Y 2004 \emph{J. Korean Phys. Soc.} \textbf{44} 672

\bibitem{Wang}
Wang J, Yang C X, Zhou P L, Jin Y D, Zhou T and Wang B H 2005
\emph{Physica} A \textbf{354} 505

\bibitem{ZhouPL}
Zhou P L, Yang C X, Zhou T, Xu M, Liu J and Wang B H 2005
\emph{New Mathematics and Natural Computation} \textbf{1} 275

\bibitem{Newman}
Newman M E J and Watts D J 1999 \emph{Phys. Rev.} E \textbf{60}
7332

\bibitem{Moelbert}
Moelbert S and Rios P De L 2002 \emph{Physica} A \textbf{303} 217

\end{references}
\end{document}